\documentclass{elsart}

\usepackage{epsfig}
\usepackage{amssymb,array,amsmath,psfrag}

\begin{document}

\begin{frontmatter}

\title{Fractality of deterministic diffusion in the nonhyperbolic climbing sine map}

\author{N. Korabel and R. Klages}

\address{Max-Planck-Institut f\"ur Physik komplexer Systeme, N\"othnitzer
Stra{\ss}e 38, D-01187 Dresden, Germany}

\begin{abstract}
The nonlinear climbing sine map is a nonhyperbolic dynamical system exhibiting
both normal and anomalous diffusion under variation of a control parameter. We
show that on a suitable coarse scale this map generates an oscillating
parameter-dependent diffusion coefficient, similarly to hyperbolic maps, whose
asymptotic functional form can be understood in terms of simple random walk
approximations. On finer scales we find fractal hierarchies of normal and
anomalous diffusive regions as functions of the control parameter. By using a
Green-Kubo formula for diffusion the origin of these different regions is
systematically traced back to strong dynamical correlations. Starting from the
equations of motion of the map these correlations are formulated in terms of
fractal generalized Takagi functions obeying generalized de Rham-type functional
recursion relations. We finally analyze the measure of the normal and
anomalous diffusive regions in the parameter space showing that in both cases
it is positive, and that for normal diffusion it increases by increasing the
parameter value.
\end{abstract}

\begin{keyword}
Deterministic diffusion \sep nonhyperbolic maps \sep fractal diffusion coefficient \sep anomalous diffusion \sep periodic windows

\PACS 05.45.Df \sep 05.10.-a \sep 05.45.Ac \sep 05.60.-k
\end{keyword}
\end{frontmatter}

\section{Introduction}

Low-dimensional time-discrete maps are among the most important models for
exploring different aspects of chaos. These systems display a very rich
dynamical behavior but are still very amenable to straightforward computer
simulations. Even more, in some cases rigorous analytical solutions are
possible. After it was realized that diffusion processes can be generated by
microscopic deterministic chaos in the equations of motion, time-discrete maps
became useful tools in deterministic transport theory. The analysis of these
simple models required to suitably combine nonequilibrium statistical
mechanics with dynamical systems theory leading to a more profound
understanding of transport in nonequilibrium situations
\cite{LL83,Ott97,Gas98,Dor99}.  However, time-discrete maps provide not only 
a suitable starting point for studying normal diffusion but also for
investigating the anomalous case
\cite{Gei82,Sch82,Gro82,Fuj82,Gro83,Tsi93,Rei94,Dan89,Pra91,Chi02,Cvi01,Art91,Tse94,Che95,Rec80,Rec81,Ish91,Zas97,Leb98,Zum93,Klaf94/95,Met00}.
Moreover, there are certain classes of more realistic models which share
specific properties of maps such as being low-dimensional and exhibiting
certain periodicities. Indeed, theoretical investigations of chaotic billiards
subject to external fields \cite{Har01}, of periodic Lorentz gases
\cite{Kla00,Kla02}, and of pendulum-like differential equations
\cite{Chi79,Hub80,Ped81,Mir85,Bla96,Fes02,Sak02} showed that many properties
of deterministic transport in maps carry over to these more complex chaotic
dynamical systems.
 
In this framework, recently a new feature of deterministic diffusion was
discovered. For simple one-dimensional hyperbolic maps it was shown that the
diffusion coefficient is typically a fractal function of control parameters
\cite{Kla95,Kla96,Kla97,Kla99}. Subsequently an analogous behavior was detected for
other transport coefficients
\cite{Gas98a,GK01}, and in more complicated models \cite{Har01,Kla00,Kla02}. 
However, up to now the fractality of transport coefficients could be assessed
for hyperbolic systems only, whereas, to our knowledge, the fractal nature of
classical transport coefficients in the broad class of nonhyperbolic systems 
was not discussed. 

On the other hand, studying nonhyperbolic dynamics appears to be more relevant
in order to connect fractal transport coefficients to some known
experiments. Here we think particularly of dissipative systems driven by
periodic forces such as Josephson junctions in the presence of microwave
radiation
\cite{Bel77,Ben81,Cir82,Mir83,Mar89,TKK02,Weiss}, superionic conductors
\cite{Bey76,Mar86}, and systems exhibiting charge-density waves \cite{Bro84}
in which certain features of deterministic diffusion were already observed
experimentally. For these systems the equations of motion are typically of the
form of some nonlinear pendulum equation. In the limiting case of strong
dissipation they can be reduced to nonhyperbolic one-dimensional time-discrete
maps sharing certain symmetries \cite{Kog83,Boh84}. The so-called climbing
sine map is a well-known example of this class of maps
\cite{Gei82,Sch82,Gro83}. 

In this paper we pursue a detailed analysis of the diffusive and dynamical
properties of the climbing sine map. Particularly, we show that the
nonhyperbolicity of this map does not destroy the fractal characteristics of
deterministic diffusive transport as they were found in hyperbolic systems. On
the contrary, fractal structures appear for normal diffusive parameters as
well as for anomalous diffusive regions. We argue that higher-order memory
effects are crucial to understand the origin of these fractal hierarchies in
this nonhyperbolic system. By using a Green-Kubo formula for diffusion, the
dynamical correlations are recovered in terms of fractal Takagi-like
functions. These functions appear as solutions of a generalized
integro-differential de Rham-type equation. We furthermore show that the
distribution of periodic windows exhibiting anomalous diffusion forms Devil's
staircase like structures as a function of the parameter and that the
complementary sets of chaotic dynamics have a positive measure in parameter
space that increases by increasing the parameter value.

Our paper is organized as follows. In Sec. II we introduce the model. In
Sec. III we explore the coarse functional form of the parameter-dependent
diffusion coefficient and discuss it in relation to previous results on
hyperbolic maps. In Sec. IV our analysis is refined revealing complex
scenarios of anomalous diffusion, which are explained in terms of correlated
random walk approximations. In Sec. V generalized fractal Takagi functions are
constructed for the climbing sine map and the connection to the diffusion
coefficient is worked out.  Periodic windows exhibiting anomalous diffusion
are studied in detail in Sec. VI. We then draw conclusions and discuss our
results in the final section.

\section{The climbing sine map}
The one-dimensional climbing sine map is defined as
\begin{figure}[htb]
\centering
\includegraphics[width=0.50\textwidth]{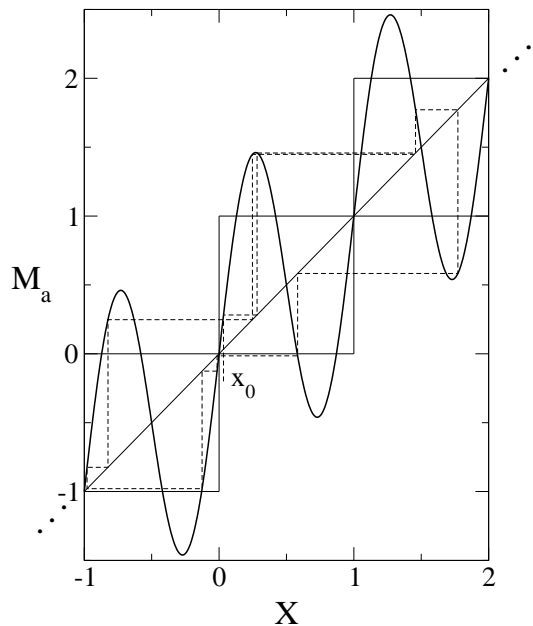}\\
\caption{Illustration of the climbing sine map for the particular parameter
value of $a=1.189$.  The dashed line indicates the orbit of a moving particle
starting from the initial position $x_0$.}%
\label{Fig1}
\end{figure}
\begin{equation}
X_{n+1} = M_a(X_n) \; , \; \; M_a (X) = X + a \sin(2 \pi X)\:,
\label{map}
\end{equation}
where $a\in\mathbb{R}$ is a control parameter and $X_n$ is the position of a
point particle at discrete time $n$. Obviously, $M_a(X)$ possesses translation
and reflection symmetry,
\begin{equation}
M_a(X + p) = M_a (X) + p \;, \; \; M_a(-X) = -M_a (X)\:.
\label{sym}
\end{equation}
\vspace*{-0.35cm}

The periodicity of the map naturally splits the phase space into different
boxes, $(p, p+1]$, $p \in \mathbb{Z}$, as shown in Fig.\
\ref{Fig1}. Eq.\ (\ref{map}) as restricted to one box, i.e., on a circle, we
call the reduced map,
\begin{equation}
m_a (x):= M_a (X) \; \mbox{mod} \; 1\:,\:x := X \; \mbox{mod} \; 1\:.
\label{reduced_map}
\end{equation}
The probability $\rho_n(x)dx$ to find a particle at a position between $x$ and
$x+dx$ at time $n$ then evolves according to the continuity equation for the
probability density $\rho_n(x)$, which is the Frobenius-Perron equation
\cite{LM94}
\begin{equation}
\rho_{n+1} (x)=\int dy \; \rho_n(y) \delta (x-m_a (y)) \:.
\label{FP_eq}
\end{equation}
The stationary solution of this equation is called the invariant density,
which we denote by $\rho^*(x)$.

Due to its nonhyperbolicity, the climbing sine map possesses a rich dynamics
consisting of chaotic diffusive motion, ballistic dynamics, and localized
orbits. Under parameter variation these different types of dynamics are highly
intertwined resulting in complicated scenarios related to the appearance of
periodic windows \cite{Sch82,Gro83}.  In order to study diffusion we will be
interested in parameters that are greater than $a > a_0=0.732644...$ for which
the extrema of the map exceed the boundaries of each box for the first time
indicating the onset of diffusive motion.

\section{Coarse structure of the parameter-dependent diffusion coefficient}

In this section we explore the relationship between nonlinear maps like the
climbing sine map and simple piecewise linear maps for which, in contrast to
the climbing sine map, the diffusion coefficient can be calculated exactly. It
is well-known that in special cases such different types of maps are linked to
each other via the concept of conjugacy. Indeed, we show that maps which are
approximately conjugate to each other exhibit a very similar oscillatory
behavior in the diffusion coefficient on coarse scales. Our argument refers to
some existing methods for calculating the diffusion coefficient of piecewise
linear maps, which we briefly review. We then describe how we numerically
calculated the complete parameter dependence of the diffusion coefficient for
the climbing sine map and discuss a first result.

\subsection{Computing and comparing the diffusion coefficient for
approximately conjugate maps} 

One speaks of normal deterministic diffusion if the mean square displacement
of an ensemble of moving particles grows linearly in time. The diffusion
coefficient is then given by the Einstein relation
\begin{equation}
D(a)=\lim_{n \rightarrow \infty} \langle X_n^2 \rangle /(2n), 
\label{D_a}
\end{equation}
where the brackets denote an ensemble average over the moving particles.

There exist various efficient numerical as well as, for some system
parameters, analytical methods to exactly compute diffusion coefficients for
piecewise linear hyperbolic maps, such as transition matrix methods based on
Markov partitions \cite{Kla95,Kla96,Kla97,Kla99}, cycle expansion
methods\cite{Cvi01,Art91,Tse94,Che95}, and more recently a very powerful
method related to kneading sequences \cite{GK01}.
 
\begin{figure}[htb]
\centering
\psfrag{Log a}{$a$}
\psfrag{Log D_rw}{$D_{rw}$}
\includegraphics[width=0.7\textwidth]{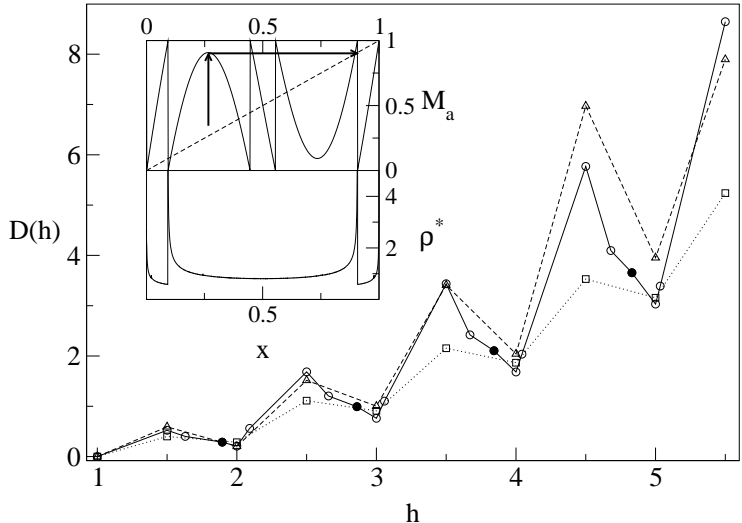}\\
\caption{Diffusion coefficient at certain Markov partition parameter values for
three different maps, which are the zig-zag map Eq.(\ref{zz_map}) (squares),
the climbing sine map (circles), and the cubic map Eq.\ (\ref{Cub_map})
(triangles). The values for the zig-zag map represent analytical results, see
Eqs.(\ref{Cub_D1},\ref{Cub_D2}), the remaining values are from computer
simulations. The lines are guides for the eyes. The inset shows an example of
a Markov partition for the climbing sine map on the unit interval and the
corresponding invariant density.}
\label{D_Mark}
\end{figure}

We first restrict our analysis of diffusion in the climbing sine map to
parameters for which there are simple Markov partitions. For one-dimensional
maps, a partition is a Markov partition if and only if parts of the partition
get mapped again onto parts of the partition, or onto unions of parts of the
partition, see Ref.\ \cite{Kla96} and further references therein.  An example
of a Markov partition consisting of five parts is shown in the inset of Fig.\
\ref{D_Mark}. In case of the climbing sine map Markov partitions can be
constructed simply by forward iteration of one of the critical points $x_c$
defined by the condition that $m_a^{'}(x_c) \equiv 0$ in the reduced map. If
higher iterations of this point fall onto a periodic orbit a Markov partition
exits. Indeed, if a Markov partition is known, for piecewise linear maps the
diffusion coefficient can often be calculated analytically via calculating the
invariant measure of the map or via computing the second largest eigenvalue of
the Frobenius-Perron operator written in form of a transition matrix.

One can now identify an infinite series of parameter values corresponding to a
certain type of Markov partition \cite{Kla95,Kla96,Kla97,Kla99}. For parameter
values which belong to such a Markov partition series the corresponding
invariant densities $\rho^{*}(x)$ have a very similar functional form. Note
that, in case of nonlinear maps, singularities in the invariant density
exactly correspond to the iteration of the critical point $x_c$ \cite{Alo96}.
An example of $\rho^{*}(x)$ for one series of parameter values (marked as
filled circles) is shown in the inset of Fig.\ \ref{D_Mark}.

By using respective series of Markov partitions piecewise linear maps can be
related to nonlinear maps.  For this purpose let us consider, along with the
climbing sine map, (i) the piecewise linear zig-zag map
\cite{Gro82,Fuj82,Art91,Tse94} 
\begin{equation}
M_{a} (x_n)= \begin{cases} 
m_1 x_n, & 0 \le x_n < b_1 \cr 
-m_2 (x_n-b_1)+a, & b_1 \le x_n < b_2 \cr
m_1(x_n-1)+1, & b_2 \le x_n < 1 
\end{cases}
\label{zz_map}
\end{equation}
with $m_1=m_2=4a-1$, $b_1+b_2=1$ and $b_1=a/m_1$, and (ii) the nonlinear cubic
map \cite{Sch82,LM94}
\begin{equation}
M_{a} (x_n)= a x^{3}_{n} - \frac{3}{2} a x_n^2 + x_n (1+\frac{1}{2}a).
\label{Cub_map}
\end{equation}
The definitions of both maps are given on the unit interval.

In order to compare the diffusion coefficient of these different maps, the
parameters $a$ were chosen such that the maps all display the same height $h$,
defined as the distance between the first iteration of the leftmost critical
point $m_a(x_c)$ and the zero bound in the first box $(0,1]$. Thus, $h=1$
corresponds to the onset of diffusion for all three maps.

For the two simple Markov partition series $m_{a}(x_c)=0$ and
$m_{a}(x_c)=0.5$, corresponding to integer and half-integer values of $h$, 
respectively, the diffusion coefficient of the zig-zag map can
be calculated analytically. For integer values of $h$ the result reads
\cite{Fuj82,Art91} 
\begin{equation}
D(h) = \frac{h(h-1)(2h-1)}{3(4h-1)}.
\label{Cub_D1}
\end{equation}
For half-integer values of $h$ the diffusion coefficient is easily calculated
analytically, e.g., by transition matrix methods \cite{Kla96,Kla99}, to
\begin{equation}
D(h)=\frac{h(4h^2-1)}{6(4h-1)}\: .
\label{Cub_D2}
\end{equation}
In case of the climbing sine map and of the cubic map the diffusion
coefficient was obtained from computer simulations by evaluating the mean
square displacement Eq.\ (\ref{D_a}) for the same series of Markov
partitions. Results are shown in Fig.\ \ref{D_Mark}. For the climbing sine map
some more Markov partition series points (alltogether five different series)
were included. For all three maps there is a very analogous oscillatory
behavior of the parameter-dependent diffusion coefficient. These oscillations
can be explained in terms of the changes of the microscopic dynamics under
parameter variation, that is, whenever there is a local maximum there is an
onset of strong backscattering in the dynamics yielding a local decrease of
the diffusion coefficient in the parameter, and vice versa at local minima
\cite{Kla95,Kla96,Kla97,Kla99}.  However, the five Markov partition series for the
climbing sine diffusion coefficient already indicate that there are more
irregularities on finer scales. For piecewise linear maps, the origin of these
irregularities was identified to be the topological instability of the
dynamics under parameter variation \cite{Kla96,Kla99}. That is, a small
deviation of the parameter changes the Markov partition and the corresponding
invariant density which, in turn, is reflected in a change of the value of the
diffusion coefficient. Note that the dependence of the diffusion coefficient
for a single Markov partition series appears to be a monotonously increasing
function of the parameter \cite{Tas}.  Nevertheless, computing $D(a)$ for more
and more Markov partitions series will reveal more and more irregularities in
$D(a)$ thus forming a fractal structure \cite{Kla95,Kla96,Kla97,Kla99}.

Since the climbing sine map shares the same topological features as piecewise
linear maps in terms of these series of Markov partitions, one may wonder
whether it is not possible to straightforwardly calculate the diffusion
coefficient for nonlinear maps from the one of piecewise linear maps by using
the concept of conjugacy \cite{Gro83,LM94,Gro77}, see also the definition in Appendix A.
 In fact, it was stated by
Grossmann and Thomae \cite{Gro83} that the diffusion coefficient is invariant
under conjugacy, however, without giving a proof. In Appendix A such a proof
is provided. Unfortunately, conjugacies are explicitly known only in very
specific cases and for maps acting on the unit interval \cite{LM94,Gro77}. As
soon as the map extrema exceed the unit interval, which is reminiscent of the
onset of diffusive behavior, only some approximate, piecewise conjugacies
could be constructed in a straightforward way, see Ref.\ \cite{Gro83} for an
example.

We now apply this reasoning along the lines of conjugacy in order to
understand the similarities between the diffusion coefficient of the three
maps as displayed in Fig.\ 2. The functional form of the cubic map can be
obtained from a Taylor series expansion of $sin(x_n)$ by keeping terms up to
third order thus representing a low-order approximation of the climbing sine
map. This seems to be reflected in the fact that at any odd integer parameter
value of $h$ the climbing sine map has an invariant density whose functional
form is very close to the one of the cubic map at parameter value $h=1$,
$\rho^*(x)=\pi^{-1}(x(1-x))^{-1/2}$. Hence, one may expect that both diffusion
coefficients are possibly trivially related to each other, however, note the
increasing deviations between the respective results at larger $h$.

For $h=1$, the cubic map and the piecewise linear zig-zag map are now in turn
conjugate to each other \cite{LM94,Gro77}. However, for $h>1$ we are not aware
of the existence of any exact conjugacy between zig-zag and cubic map. Still,
along the lines of Ref.\ \cite{Gro83} one can at least approximately relate
both maps to each other via using piecewise conjugacies. This explains why the
zig-zag map and the climbing sine map display qualitatively the very same
oscillatory behavior in the diffusion coefficient, somewhat linked by
diffusion in the cubic map.

In summary, by using Markov partitions and by arguing with the concept of
conjugacy we have shown that the structure of the diffusion coefficient for
the nonlinear climbing sine map has much in common with the one of respective
piecewise linear maps, in the sense of displaying a non-trivial oscillatory
parameter dependence.  However, to use conjugacies in order to exactly
calculate the diffusion coefficient for nonlinear maps does not appear to be
straightforward \cite{Alo96}, hence in the following we restrict ourselves to
alternative methods as discussed in the next subsection.

\subsection{Complete parameter dependence of the climbing sine diffusion coefficient}

In order to obtain the full parameter dependence for the diffusion coefficient
of the climbing sine map we numerically evaluated the Green-Kubo formula for
diffusion in maps
\cite{Gas98,Dor99,Fuj82,Kla02,Kla96,Gas98a} reading
\begin{equation}
D_n(a) = \langle j_a(x) J_a^n(x) \rangle -
\frac{1}{2}\langle j_a^{2}(x) \rangle \:.
\label{GK}
\end{equation}
Here the angular brackets denote an average over the invariant density of the
reduced map, $\langle \ldots \rangle = \int d x \rho^* (x)\ldots$. The jump
velocity $j_a(x)$ is defined by
\begin{equation}
j_a(x_n):=[X_{n+1}]-[X_n]\equiv [M_a(x_n)] \:,
\label{Jumps_Vel}
\end{equation}
where the square brackets denote the largest integer less than the argument. 
\begin{figure}[htb]
\centering
\psfrag{b}{$a$}
\includegraphics[width=0.7\textwidth]{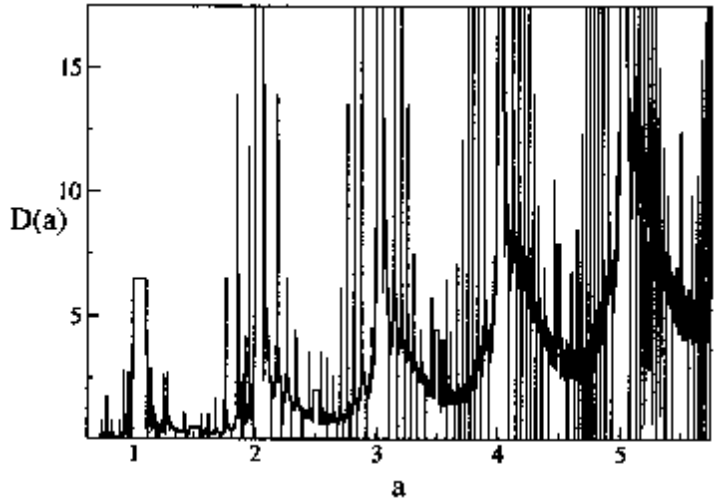}\\
\caption{Diffusion coefficient for the climbing sine map over a large range of parameter
values. Note the oscillations on large and small scales. The small scale
fluctuations represent regions of anomalous diffusion where the diffusion
coefficient either diverges or vanishes. Some of the divergent regions are cut
off after a certain number of iterations showing plateaus instead of
singularities. The data set consists of $265005$ points.}
\label{D_5box}
\end{figure}
The sum
\begin{equation}
J_a^n(x) = \sum_{k=0}^n j_a (x_k) 
\label{Vel_func}
\end{equation}
gives the integer value of the displacement of a particle after $n$ time steps
that started at some initial position $x$, and we call it jump velocity
function. Eq.\ (\ref{GK}) defines a time-dependent diffusion coefficient
which, in case of normal diffusion, converges to
\begin{equation}
D(a)=\lim_{n\to\infty}D_n(a) \:. 
\label{Dlim}
\end{equation}
In our simulations for Fig.\ \ref{D_5box} we truncated $J_a^n(x)$ after $7$ time steps. 
The invariant density was obtained by solving
the continuity equation for $\rho^* (x)$ Eq.\ (\ref{FP_eq}) with the histogram
method of Ref.\ \cite{LL83}. Note that both the integer displacement and the
density are coupled via Eq.\ (\ref{GK}).  Results for $D(a)$ are shown in
Fig.\ \ref{D_5box} for a large range of parameters demonstrating a highly
non-trivial behavior of the diffusion coefficient.  The large-scale
oscillations as predicted from the simple Markov partition series (see Fig.\
\ref{D_Mark}) are still clearly seen, however, on top of this there exist
further oscillations on finer scales. These are regions of anomalous diffusion
that manifest themselves in form of abrupt divergences, $D(a)
\rightarrow \infty$, or by a vanishing diffusion coefficient, $D(a)
\rightarrow 0$. 

\section{Simple and correlated random walk approximations}

In this section we study the parameter-dependent diffusion coefficient in more
detail. Based on the Green-Kubo formula we derive a systematic hierarchy of
approximations for the diffusion coefficient and show how they can be used to
understand the complex behavior of this curve in more detail.

\subsection{Asymptotic functional form of the diffusion coefficient on large
scales} 

We are first interested in understanding the coarse functional form of the
parameter-dependent diffusion coefficient in the limit of very small and very
large parameter values. For this purpose we use simple random walk
approximations that are based on the assumption of a complete loss of memory
between the single jumps. Such an analysis was already performed for
hyperbolic piecewise linear maps \cite{Kla96,Kla97}. Here we apply the same
reasoning to the nonlinear case of the climbing sine map.

We start in the limit of very small parameter values, i.e., near the onset of
diffusion. Here we assume that particles make either a step of length one to
the left or to the right, or just remain in the box. The transition
probability is then given by integrating over the respective invariant density
in the escape region. Putting all this information into Eq. (5) yields
\cite{Sch82}
\begin{equation}
D(a) \simeq \rho (x_c) (2\epsilon /a_0 \pi^2)^{1/2},
\label{RW}
\end{equation}
where $\epsilon=a-a_0$. Making the additional approximation that $\rho (x)
\simeq 1$ we get
\begin{equation}
D(a) \simeq 0.525 \epsilon ^{1/2}, \; \; \; \epsilon \ll 1.
\label{RW_1}
\end{equation}
The other limiting case concerns values of $a\gg 1$. Here the precise value of
the width of the escape region is much less important than the precise value
of the step length which is very large, hence by again assuming that $\rho (x)
\simeq 1$ Eq. (5) can be approximated to \cite{Kla96,Kla97}
\begin{equation}
D(a) \approx \frac{1}{2}\int_{0}^{1} dx \; (M_a(x)-x)^2 \approx \frac{a^2}{4}, \; \; \; a \gg 1.
\label{RW_2}
\end{equation}
\begin{figure}[htb]
\centering
\psfrag{Log a}{$a$}
\psfrag{Log D}{$D_{rw}$}
\includegraphics[width=0.7\textwidth]{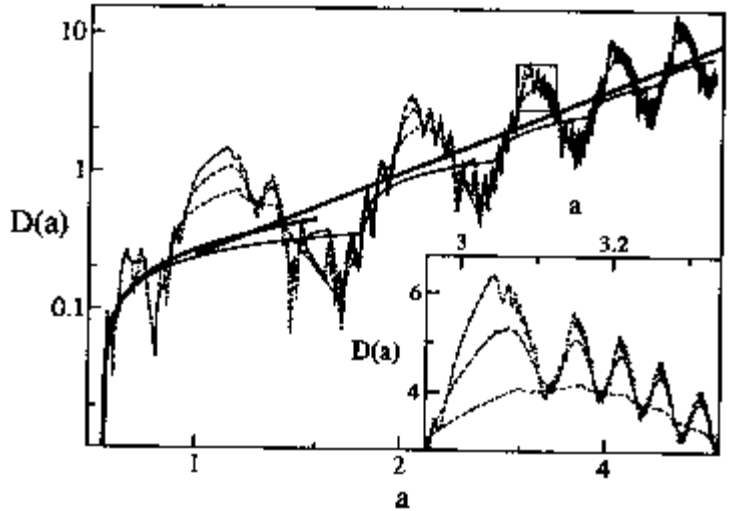}\\
\caption{Dynamical crossover in the climbing sine map. The two asymptotic regimes at
small and large parameter values are approximated by simple random walk
arguments yielding the functional forms shown as bold lines. The series of
more irregular curves corresponds to higher-order correlated random walk
approximations with $n=0,1,2,3$, which give an indication of the exact functional form.  The
magnification (inset in normal scale) of a small region around $a=3$ shows
more irregularities on a finer scale pointing towards a fractal structure of
the diffusion coefficient.}%
\label{Crisis}
\end{figure}
These two asymptotic random walk approximations are shown in Fig.\
\ref{Crisis} as bold lines. One can clearly see that there is a
dynamical crossover between the different functional forms of these two
asymptotic regimes. This crossover was first observed in piecewise linear maps
and appears to be typical for diffusive systems exhibiting some spatial
periodicity \cite{Kla96,Kla97}. It was lateron also verified for diffusion in
the periodic Lorentz gas \cite{Kla00}. The coarse functional form of the
random walk approximations should be compared to a certain series of
higher-order approximations based on the Green-Kubo formula Eq.\ (\ref{GK}),
which is also shown in the figure. These refined approximations are getting
closer and closer to the exact functional form, as explained below, and thus
give a good indication for these exact values.

\subsection{Fine scale of the diffusion coeffcient: anomalous diffusion and bifurcations}

In the previous subsection we have outlined two simple random walk
approximations for diffusion that do not include any memory effects. However,
one can do better by systematically evaluating the single terms as contained in
the series expansion of the Green-Kubo formula Eqs.\
(\ref{GK},\ref{Vel_func}). For a simple piecewise linear map and for the
periodic Lorentz gas this was done in Ref.\ \cite{Kla02} and provided a
simple approach to understand the origin of complex structures in the
diffusion coefficient on fine scales.

The basic idea of this approach is as follows: The Green-Kubo formula Eq.\
(\ref{GK}) splits the dynamics into an inter-box dynamics, in terms of integer
jumps, and into an intra-box dynamics, as represented by the invariant
density. We first approximate the invariant density in Eq.\ (\ref{GK}) to
$\rho (\tilde{x})\simeq 1$ irrespective of the fact that it is typically a
very complicated function of $x$ and $a$
\cite{Sch82,Kla96}. The resulting approximate diffusion coefficient  we label with a
superscript in Eq.\ (\ref{GK}), $D_n^1(a)$. The term for $n=0$ obviously
excludes any higher-order correlations and was already worked out in form of
the simple random walk approximation Eqs.\ (\ref{RW})-(\ref{RW_2}).

The generalization $D_n^1(a)\;,\;n>0$, which systematically includes more and
more dynamical correlations, may consequently be denoted as {\em correlated
random walk approximation} \cite{Kla02}. We now use this expansion to analyze
the parameter dependence of the diffusion coefficient of the climbing sine map
in terms of such higher-order correlations.
\begin{figure}[htb]
\centering
\psfrag{D^n_b}{$D_a^n$}
\psfrag{b}{$a$}
\psfrag{(a)}{$(a)$}
\psfrag{(b)}{$(b)$}
\psfrag{(c)}{$(c)$}
\includegraphics[width=0.7\textwidth]{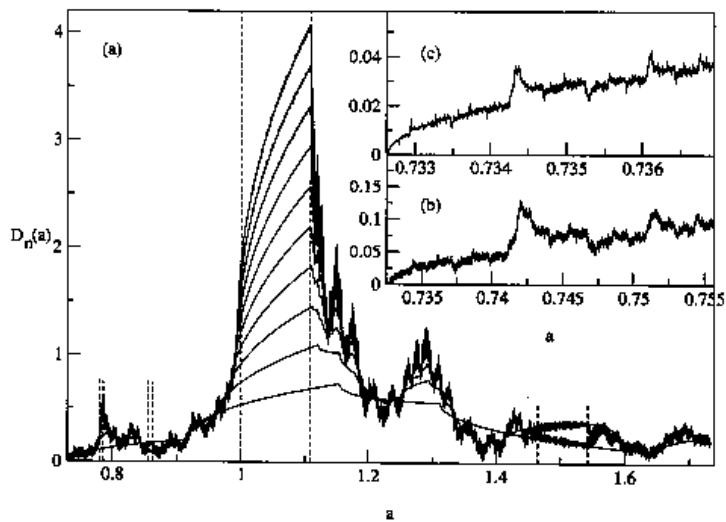}\\
\caption{(a) Sequence of correlated random walks $D_n^1(a)$, see Eq.\ (\ref{GK})
with uniform invariant density $\rho(x)\simeq 1$, for $n=1, \cdots, 10$. Note
the quick convergence for normal diffusive parameters. The dashed lines define
periodic windows, which are the same as in Fig.\ref{Dif}. The insets (b) and
(c) contain two magnifications of (a) in the region close to the onset of
diffusion for $D_{10}^1(a)$ only. They show self-similar behavior on smaller
and smaller scales.}%
\label{FRW}
\end{figure}
\noindent
Fig.\ \ref{Crisis} depicts results for $D_n(a)$ at $n=0,1,2,3$ 
over a large range of parameters, whereas Fig.\ \ref{FRW} (a) presents a
respective detailed analysis for the region close to the onset of diffusion,
i.e., for parameters $a \in (0.732644, 1.742726]$, showing results
for $D_n(a)$ at $n=1,\ldots,10$. The series of approximations in Fig.\
\ref{Crisis} clearly reveals finer and finer sequences of oscillations that
eventually converge to a fractal structure, as is particularly shown in the
inset of this figure. However, here the order of the expansion is not large
enough to identify parameter regions of anomalous diffusion.  These regions
can be better seen in Fig.\ \ref{FRW}, where three different cases of
parameter regions can be distinguished: (i) regions with quick convergence of
this approximation corresponding to normal diffusion (ii) divergence of
$D_n^1(a)$ corresponding to ballistic motion, in agreement with
$D(a)\to\infty$, and (iii) localized dynamics where $D_n^1(a)$ alternates in
$n$ between two solutions, with $D(a)\to0$ for the exact diffusion
coefficient. This oscillation points to the dynamical origin of localization
in terms of certain period-two orbits. That these approximate solutions are
non-zero is due to the fact that the invariant density was set equal to
one. The dashed lines in Fig.\
\ref{FRW} indicate the largest regions of anomalous diffusion. 
The approximate diffusion coefficient $D_{10}^1(a)$ of this figure is compared
to the ``numerically exact'' one in Fig.\ \ref{Dif}. Here ``numerically
exact'' we wish to be understood in the sense that no further ad
hoc-approximations are involved, i.e., we evaluated the Green-Kubo formula
according to the numerical method described in Sec.\ III. by truncating it after 
$20$ time steps. This comparison
shows that in case of normal diffusion our approximation nicely reproduces the
irregularities in the non-approximated diffusion coefficient. Like the inset
of Fig.\ \ref{Crisis}, the magnifications in Fig.\ref{FRW} give clear evidence
for a self-similar structure of the diffusion coefficient. These results thus
show that dealing with correlated jumps only yields a qualitative and to quite
some extent even quantitative understanding of the regions of normal and
anomalous diffusion in the climbing sine map.

The impact of specific features of the microscopic dynamics on the diffusion
coefficient is nicely elucidated by comparing the bifurcation diagram of the
reduced climbing sine map Eq.\ (\ref{reduced_map}) with the numerically exact
diffusion coefficient, see Fig.\ \ref{Dif}.  As one can see in the upper panel
of Fig.\ \ref{Dif}, the bifurcation diagram consists of (infinitely) many
periodic windows.  Whenever there is a window the dynamics of Eq.\ (\ref{map})
is either ballistic or localized \cite{Sch82,Fuj82}. Fig.\
\ref{Dif} demonstrates the strong impact of this bifurcation scenario on the
diffusion coefficient. For localized dynamics, orbits are confined within some
finite interval in phase space implying subdiffusive behavior for which the
diffusion coefficient vanishes, whereas for ballistic motion particles
propagate superdiffusively with a diverging diffusion coefficient being
proportional to $n $. Only for normal diffusion $D(a)$ is nonzero, finite and
the limit in Eq.\ (\ref{Dlim}) exists. At the boundaries of each periodic
window the diffusion coefficient is related to intermittent-like transient
behavior eventually resulting in normal diffusion with $D(a)\sim a^{(\pm
\frac{1}{2})}$ \cite{Gei82,Sch82,Gro83,Tsi93}.
\begin{figure}[htb]
\centering
\psfrag{x}{$x$}
\psfrag{b}{$a$}
\psfrag{D_b}{$D_a$}
\includegraphics[width=0.7\textwidth]{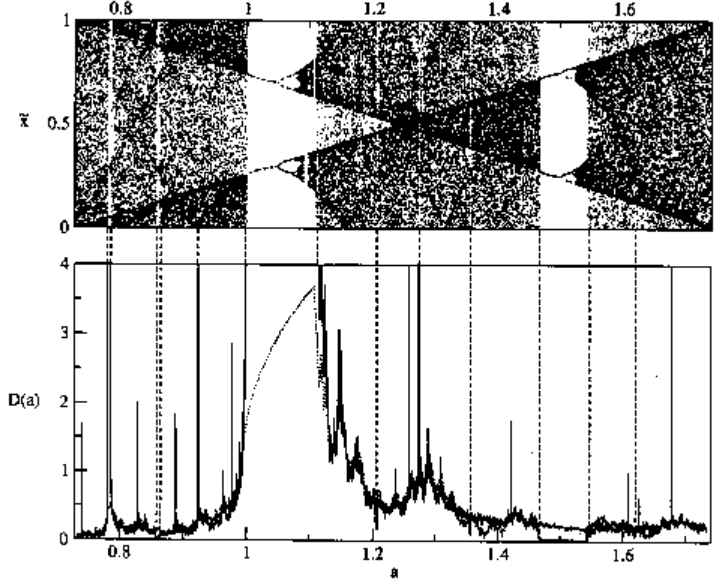}\\
\caption{Upper panel: bifurcation diagram for the climbing sine map.
Lower panel: diffusion coefficient computed from simulations as a function of
the control parameter $a$ in comparison with the correlated random walk
approximation $D_{10}^1(a)$ (dots). The dashed vertical lines connect regions
of anomalous diffusion, $D(a)\to\infty$ or $D(a) \to 0$, with 'ballistic' and
'localized' windows, respectively, of the bifurcation diagram.}%
\label{Dif}
\end{figure}

\section{Diffusion coefficient in terms of fractal generalized Takagi functions}

In this section we further analyze the dynamical origin of the different
structures in the parameter-dependent diffusion coefficient by constructing
objects called fractal generalized Takagi functions. These functions somewhat
resemble usual Takagi functions but, as will be shown, they fulfill a more
complicated type of functional recursion relations than standard de Rham-type
equations. Interestingly, Takagi functions were known to mathematicians since
about a hundred years \cite{Tak73,dRh57,Hat84}, however, in the field of
chaotic transport they were appreciated by physicists only very recently
\cite{Gas98,Dor99,Kla96,Tas93,Gas01}.

We first show how to construct fractal generalized Takagi functions and study
their properties with respect to the three different types of dynamics in the
climbing sine map. We then relate these objects directly to the diffusion
coefficient.

From the definition of the time-dependent jump velocity function there follows
the recursion relation \cite{Kla95,Kla96}
\begin{equation}
J_a^n (x) =  j_a(x) + J_a^{n-1} (m_a(x)).
\label{rec}
\end{equation}
Since the time-dependent jump velocity function $J_a^n$ is getting extremely
complicated after some time steps, we introduce the more well-behaved function
\begin{equation}
T_a^n(x) := \int_0^x J_a^n (z) \; dz, \; \; \; T_a^n(0)\equiv
T_a^n(1) \equiv 0.
\label{Takagi}
\end{equation}
Integration of Eq.\ (\ref{rec}) yields the recursive functional equation
\begin{equation}
T_a^{n}(x) = t_a (x) + \frac{1}{m^{'}_{a}(x)} \;
T_a^{n-1} (m_a(x)) - I(x)
\label{TakEq}
\end{equation}
with the integral term
\begin{equation}
I(x) = \int_{0}^{m_a(x)} dz g_a''(z) T_a^{n-1} (z)\;,
\label{TakEq2}
\end{equation}  
\begin{figure}[htb]
\centering
\includegraphics[width=0.55\textwidth]{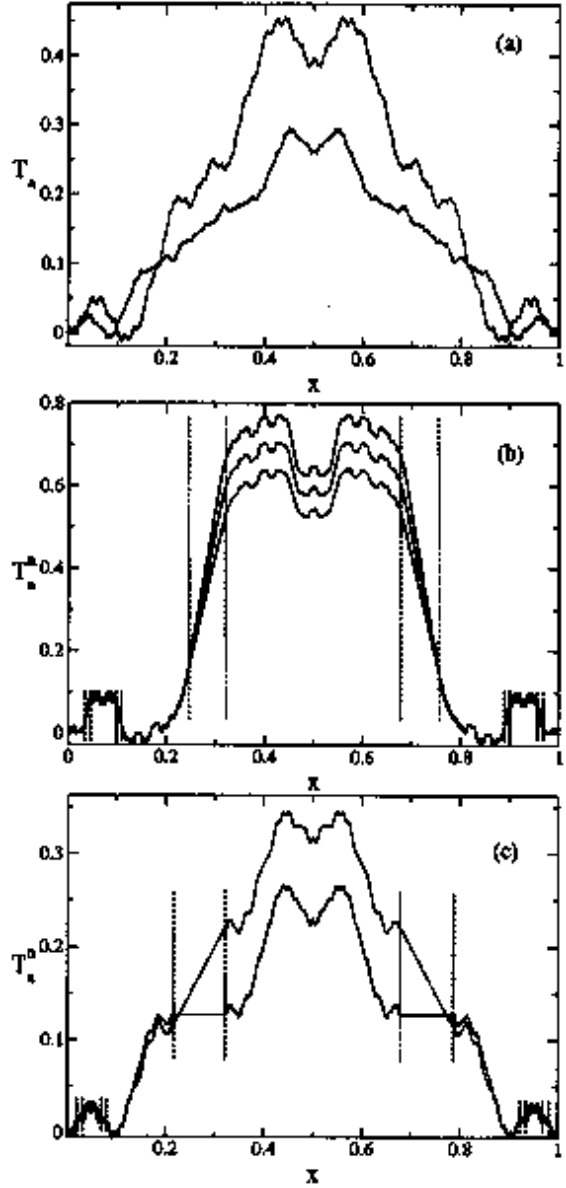}\\
\caption{(a): Generalized fractal Takagi functions 
for the diffusive climbing sine map with parameters $a=1.2397$ (upper curve)
and $a=1.7427$ (lower curve).  (b), (c): An example of nonconverging
iterations of the generalized fractal Takagi functions for the climbing sine
map with parameters corresponding to (b) ballistic dynamics at $a=1.0$ and to
(c) localized dynamics at $a=1.5$, both for the time steps $n=5, 6, 7$. Note
the divergence of the iterations in (b) and the alternation between two states
in (c).}%
\label{TakFIG}
\end{figure}
\noindent 
where $g_a''(z)$ is the second derivative of the inverse function of $m_a(x)$.
The function $t_a (x)$ is given by
\begin{equation}
t_a (x) = \int dz j_a (z) = x j_a(x) + c(x)\;,
\label{t_a}
\end{equation}  
where $c(x)$ is defined to be constant on each subinterval where the jump
velocity $j_a(x)$ has a given value. This constant is fixed by the condition
for $t_a(x)$ to be continuous on the unit interval supplemented by
\begin{equation} 
t_a(0)=t_a(1) \equiv 0.
\label{t_0}
\end{equation}
The generalized Takagi function is now defined in the long-time limit of Eq.\
(\ref{TakEq}),
\begin{equation}
T_a(x)=\lim_{n \to\infty} T_a^n(x)\;.
\label{TakFun}
\end{equation} 
For piecewise linear hyperbolic maps $I(x)$ in Eq.\ (\ref{TakEq2}) simply
disappears, and the derivative in front of the second term of Eq.\ (\ref{TakEq})
reduces to the local slope of the map thus recovering ordinary de Rham-type
equations \cite{Gas98,Dor99,Kla96,Gas98a,GK01}. It should be noted that for
smooth nonlinear maps like the climbing sine map the reduced map $m_a(x)$ is
generally not invertible.  In order to define a local inverse of $m_a(x)$, we
split the unit interval into subintervals on which this function is piecewise
invertible.  Thus Eq.\ (\ref{TakEq}) should be understood as a series of
equations where each part is defined for a respective piecewise invertible
part of $m_a(x)$. The detailed derivation of Eq.\ (\ref{TakEq}) is given in
Appendix B.

It is not known to us how to directly solve this generalized de Rham-type
integro-differential equation for the climbing sine map, however, solutions
can alternatively be constructed from Eqs.\ (\ref{Takagi}), (\ref{t_a}),
(\ref{t_0}) starting from simulations. Results are shown in Fig.\
\ref{TakFIG}.  For normal diffusive parameters the limit of Eq.\
(\ref{TakFun}) exists and the respective curves are fractal on the whole unit
interval somewhat resembling standard Takagi functions
\cite{Gas98,Dor99,Kla96,Tas93,Gas01}. However, in case of periodic windows 
$T_a^n(x)$ either diverges due to ballistic flights, or it oscillates
indicating localization. Interestingly, in these functions the corresponding
attracting sets appear in form of smooth, non-fractal regions on fine scales
as marked by the dashed lines in Fig.\ \ref{TakFIG}.

The diffusion coefficient can now be formulated in terms of these fractal
functions by carrying out the integrations contained in Eq.\ (\ref{GK}).  For
simplicity we restrict ourselves to the parameter region of $a \in (0.732644,
1.742726]$ in which the respective solution reads
\begin{equation}
D(a)= 2 \left[ T_a(x_2)\rho (x_2)-T_a(x_1)\rho (x_1) \right]
- D_0^{\rho}(a),
\label{DTak}
\end{equation}
where $x_i,\;i=1,2$, is defined by $[M_a(x_i)]:=1$, and
$D_0^{\rho}(a):=\int_{x_1}^{x_2} d x \rho (x)$. Our previous approximation
$D_n^1(a)$ with $\rho (x) \simeq 1$ is recovered from this equation in form of
\begin{equation}
D_n^1(a)=2 \left[ T_a(x_2)-T_a(x_1) \right] - D_0^1(a)\;.
\label{DTak2}
\end{equation}
Hence, Eqs.\ (\ref{DTak}), (\ref{DTak2}) explicitly relate the generalized
fractal Takagi functions shown in Fig.\ \ref{TakFIG} to the fractals of Fig.\
\ref{Dif}.

\section{Periodic windows}

One of the most important problems regarding periodic windows in maps remains
the question of their total measure. Much understanding has been achieved for
one-dimensional unimodal maps \cite{Sha64,Met73,Jac78,Fei79/79a/83,Yor85,Far85,Gre83}.  Based on the
Sharkovskii theorem about the ordering of periodic orbits \cite{Sha64},
Metropolis, Stein and Stein organized periodic windows in universal symbolic
sequences (U-sequences) such that the sequence of next order is uniquely
determined by the previous one \cite{Met73}. Later Jacobson came up with the
proof that chaotic parameter values in one-dimensional unimodal maps with a
single maximum do have positive measure \cite{Jac78}. Related numerical
studies were made by Farmer \cite{Far85}. Furthermore, it was shown that
periodic windows in such a map form so-called fat fractal Cantor-like sets
with positive measure.

However, for diffusive maps on a line, apart from the preliminary studies of
Refs.\cite{Sch82,Fuj82}, nothing appears to be known. On the other hand, as
was exemplified in Sec. IV there is an intimate relation between the irregular
structures of the diffusion coefficient and the occurrence of periodic
windows. Hence, in this section we investigate the periodic windows for the
climbing sine map in full detail.

Due to the spatial extension of our model a new type of periodic motion, which
is not present in unimodal maps, exists, which is that particles move on
average in one direction. These {\it ballistic solutions} are born through
tangent bifurcations, further undergo a Feigenbaum-type scenario and die at
crises points \cite{Sch82,Fuj82}.  {\it Localized solutions}
occurr at even periods only and start with tangent bifurcations followed by a
symmetry breaking at slope-type bifurcation points
\cite{Sch82,Fuj82}.  In this case the bifurcation scenario is
much more complex.  Obviously, periodic windows are related to certain
periodic orbits, thus there are infinitely many of them, and they are believed
to be dense in the parameter space \cite{Ott97}.
\begin{figure}[htb]
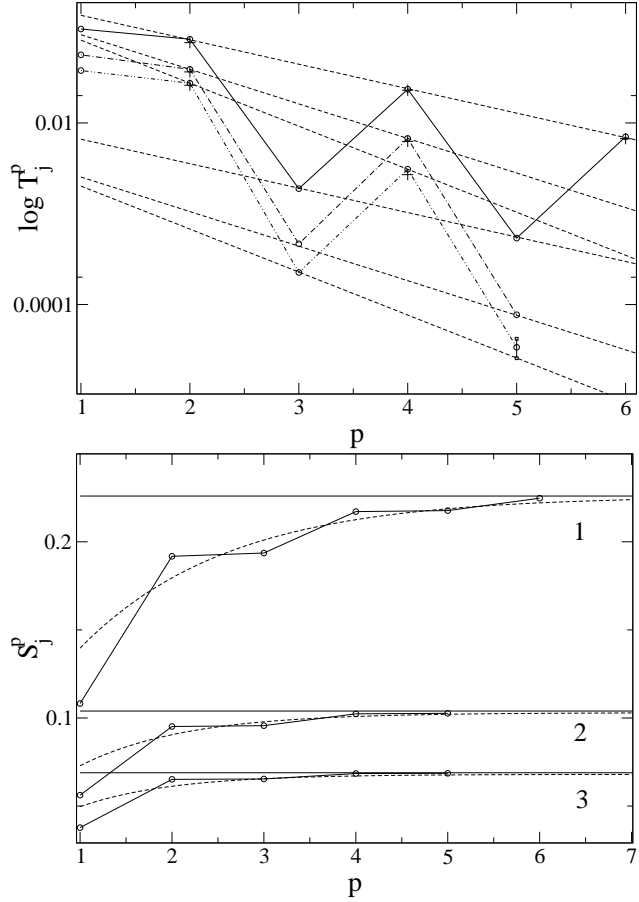

\centering
\includegraphics[width=0.6\textwidth]{Fig8_1.eps}\\
\includegraphics[width=0.6\textwidth]{Fig8_2.eps}\\
\caption{Upper panel: The total measure $T^p_j$ of all
period $p$-windows (lines with symbols) for the first three parameter
intervals (from top to bottom) as defined in the text. The dotted
lines represent exponential fits; for the parameters see Table \ref{fit}. The
measures corresponding to windows with localized orbits are shown as pluses.
Lower panel: The partial sum $S_{j}^p$ for all periodic windows at a certain
period $p$. The dashed curves represent approximations as calculated from
Eq. (\ref{Sp}), the straight lines are their limiting values at $p \rightarrow
\infty$.}%
\label{Fig5}
\end{figure}

By dividing the parameter line into subsets labeled by the integer value of
the map maximum on the unit interval, $\left[ M_a(X_{cr}) \right] = j, \; j
\in\mathbb{Z}$, we computed all windows of a certain
period $p$ in a certain subset. The numerical procedure which was used for
these computations is outlined in Appendix C.  Let $T_{j}^p$ denote the total
measure of all period $p$-windows in a subset $j$ and let $S_{j}^p$ be the
partial sum of $T_{j}^p$ defined by $S_{j}^p =\sum_{i=1}^{p} T_{j}^i$.  In
Fig. \ref{Fig5} $\log T_{j}^p$ is plotted as a function of the period for the
three first subsets $j=0,1,2$. The measures corresponding to windows with
localized orbits are shown in Fig. \ref{Fig5} as pluses. Is it clear that they
make the major contribution to the total measure for even periods hence
explaining the origin of the pronounced oscillatory behavior of $T^j_p$.
\begin{table}[htb]
\centering
\caption{Fit parameters for the exponential decrease of the measure at even and odd periods for 
the first three subsets of the map control parameter labeled by
$j$. \label{fit}}
\begin{tabular}{lllll}
$j$ & $a_{j}^{even}$ & $b_{j}^{even}$ & $a_{j}^{odd}$ & $b_{j}^{odd}$ \\
\hline
$0$ & $0.284$ & $0.61 \pm 0.01$ & $0.012$ & $0.62$ \\
\hline
$1$ & $0.224$ & $0.87 \pm 0.01$ & $0.007$ & $0.89$ \\
\hline
$2$ & $0.242$ & $1.01 \pm 0.02$ & $0.006$ & $0.98$
\end{tabular}
\end{table}
\begin{figure}[htb]
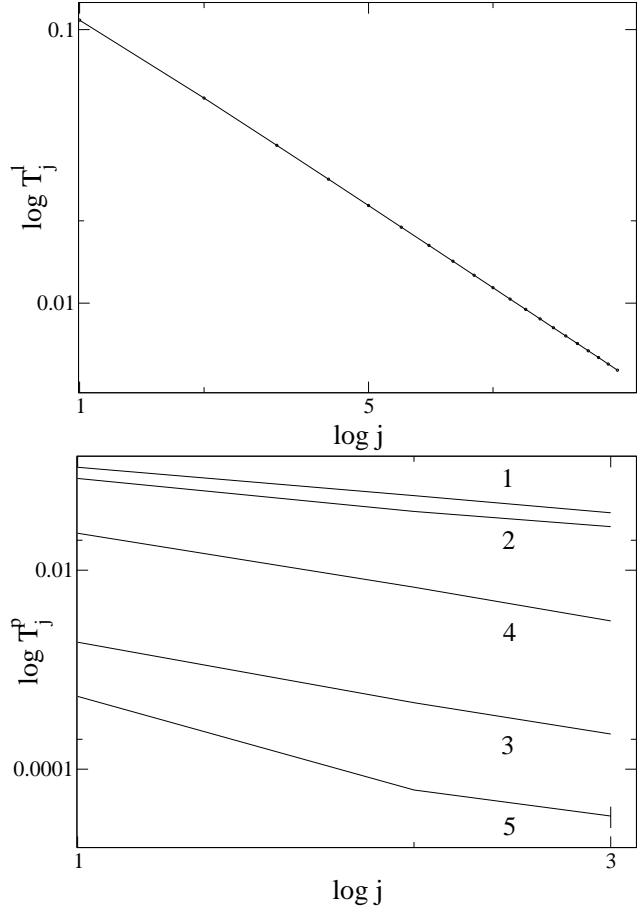

\centering
\includegraphics[width=0.6\textwidth]{Fig9_1.eps}\\
\includegraphics[width=0.6\textwidth]{Fig9_2.eps}\\
\caption{Upper panel: Total measure $T^1_j$ for period one-windows as a
function of the control parameter interval labeled by $j$. The slope of the
line appears to be $-1$ (see text).  Lower panel: the same as on the upper
panel but here for different window periodicities $p$, and for smaller
$j$. The labels at the single graphs give the value of $p$.}
\label{Fig6}
\end{figure}

For odd periods all windows are due to ballistic orbits. Thus, we more
carefully distinguish between even and odd periods. The dotted lines in Fig.\
\ref{Fig5} represent exponential fits to the functional dependence of the measure at
even and odd periods according to $a^{even}_j \exp (- b^{even}_j p)$ and
$a^{odd}_j \exp (- b^{odd}_j p)$, where $j$ stands for the box number
$j=0,1,2$, as defined above. The fit parameters $a_j, b_j$ are given in Table
\ref{fit}.  From Fig.\ \ref{Fig5} and Table \ref{fit} one can conclude that
the total measures at even and odd periods at a certain label $j$ decrease
approximately with the same rate, $b_{j}^{even} \sim b_{j}^{odd} \simeq b_j$. The
exponential decrease of $T^p_j$ suggests that the measure of chaotic solutions
in each box, which is complementary to the measure of periodic windows, is
indeed positive. Based on the information of $T^p_j$, it is straightforward to
approximate the total measure of all the periodic windows in the $j$th box by
\begin{equation}
S_j =  T^1_j+ {\sum_{i=2}^{\infty}}^{\prime} T^i_j + {\sum_{i=3}^{\infty}}^{\prime \prime} T^i_j  
\simeq T^1_j + \frac{a^{even}_j+a^{odd}_j e^{-b_j}}{e^{2 b_j} - 1}
\label{Sj}
\end{equation}
and to approximate the parameter dependence of the partial sum $S^p_j$ to 
\begin{equation}
S^p_j \simeq T_j^1 e^{-b_j (p-\frac{1}{2})} + S_j (1 - e^{- b_j
(p-\frac{1}{2})})\;.
\label{Sp}
\end{equation}
Sums with one or two primes go only over even or odd terms, respectively.  In
the lower panel of Fig.\ \ref{Fig5} results for $S_j$ are shown as horizontal
lines, the dashed lines in the upper panel (without symbols) are the
approximations for $S^p_j$ according to Eq.\ (\ref{Sp}).

The values for the measure of all periodic windows in the $j$th box, $S_j$,
and the measure of the chaotic solutions $C_j = \Delta_j - S_j$, where
$\Delta_j$ is the total measure of the $j$ box, are listed in Table
\ref{mes} for the first three subsets of the control parameter. 
\begin{table}
\centering
\caption{The Lebesque measure $\Delta_j$, the total measure of periodic
windows $S_j$, and the complementary measure of chaotic solutions $C_j$ for
the first three subsets of the map control parameter labeled by
$j$. \label{mes}}
\begin{tabular}{llll} 
$j$ & $\Delta_j$ & $S_j$ & $C_j$ \\
\hline
$0$ & $1.01008$ & $0.226 \pm 0.002$  & $0.783 \pm 0.002$ \\
\hline
$1$ & $1.00265$ & $0.103 \pm 0.002$ & $0.898 \pm 0.002$ \\
\hline
$2$ & $1.00123$ & $0.068 \pm 0.002$ & $0.932 \pm 0.002$
\end{tabular}
\end{table}
\begin{figure}[htb]
\centering
\includegraphics[width=0.7\textwidth]{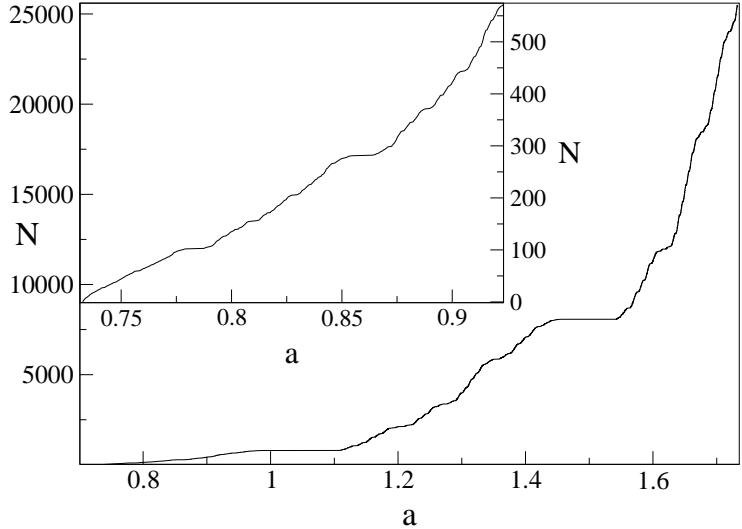}\\
\caption{Devil's staircase like structure formed by the distribution of
periodic windows as a function of the control parameter.  $N$ is the
integrated number of period six-windows.  The inset shows a blowup of the
initial region.}%
\label{Fig7}
\end{figure}

The main message of Table\ \ref{mes} is that the measure of periodic windows
is different for a different subset $j$ and obviously decreases by increasing
the control parameter. Correspondingly, the measure of chaotic solutions
increases as the parameter of the system is getting larger. In order to make
this more quantitative we consider the dependence of the measure of period
$p$-windows as a function of the box index $j$.  First we check period
one-windows. Since in each box there exists only one window of this period, we
are able to go to up to $j=20$.  In the upper panel of Fig.\ \ref{Fig6} the
logarithm of $T^1_j$ is plotted against $\log j$. We find that the slope of
this function almost exactly equal to $-1$. The behavior of $T^p_j$
for different $p$ is shown in the lower panel of Fig.\ \ref{Fig6}. The slope
of the line for period two is also $-1$, for period three and four it is $-2$,
and for period five it is $-3$.  Fig.\ \ref{Fig6} shows that even and odd
periods decrease with respectively different laws, where the decay rate
appears to be precisely given by the periodicity of the windows according to
$1/j^{p/2}$ for even periods and $1/j^{(p+1)/2}$ for odd periods.
\begin{figure}[htb]
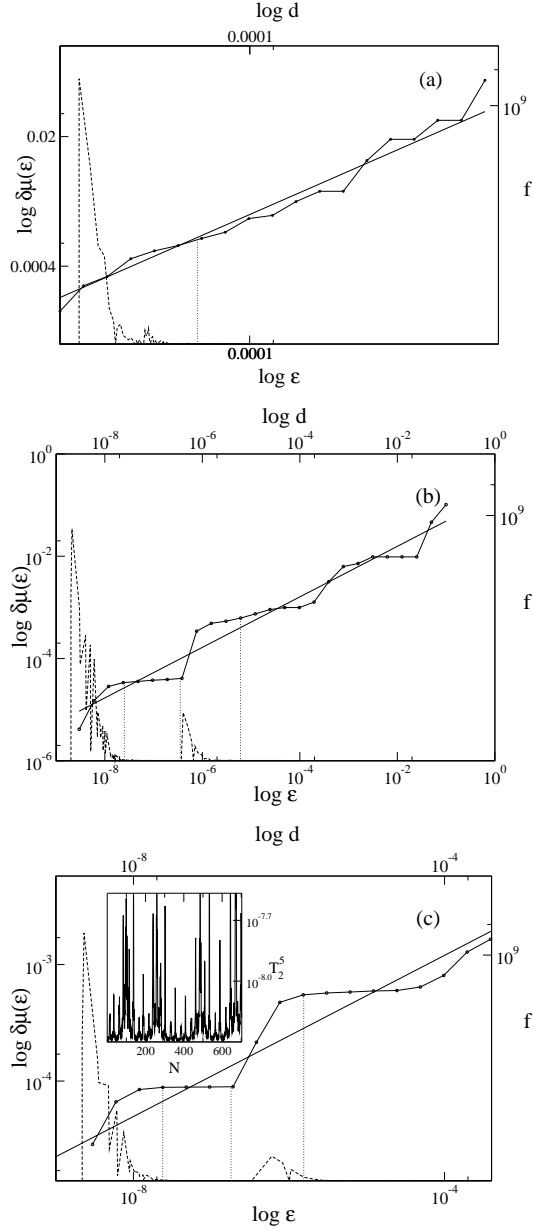

\centering
\includegraphics[width=0.5\textwidth]{Fig11_1.eps}\\
\vspace*{0.2cm}
\includegraphics[width=0.5\textwidth]{Fig11_2.eps}\\
\vspace*{0.2cm}
\includegraphics[width=0.5\textwidth]{Fig11_3.eps}\\
\caption{Difference $\delta \mu=\mu(\varepsilon)-\mu(0)$, where $\mu(\varepsilon)$
is the total measure of windows that are smaller than $\varepsilon$. 
Results are plotted for the first three subsets $j=0, 1, 2$ of the map
control parameter, from top to bottom. In all cases, the slope of the solid
lines is approximately $0.45$.  The dashed lines in each graph represent the
corresponding coarse-grained window distribution functions. In the inset of
Fig.\ (c) the measure of any single period five-window in the subset $j=2$ is
shown with respect to an integer label that accounts for the ordering
according to the map control parameter.}%
\label{Fig8}
\end{figure}

In order to analyze the structure of the regions of anomalous diffusion in the
parameter space, we sum up the number of period six-windows as a function of
the parameter, that is, the total number is increased by one for any parameter
value at which a new period six-window appears. This sum forms a Devil's
staircase like structure in parameter space indicating an underlying Cantor
set like distribution for the corresponding anomalous diffusive region, see
Fig.\ \ref{Fig6}. Since the Lebesque measure of periodic windows is positive,
this set must be a fat fractal \cite{Yor85}. Its self-similar structure can
quantitatively be assessed by computing the so-called fatness exponent.
Following \cite{Far85}, let $h(\varepsilon)$ be the total measure of all
periodic windows whose width is greater than or equal to $\varepsilon$. Define
the coarse-grained measure as $\mu(\varepsilon) = \Delta - h(\varepsilon)$,
where $\Delta$ is the total measure of a box related to the control
parameter. For quadratic maps on the interval, it was conjectured and
confirmed numerically that $\mu(\varepsilon)$ asymptotically scales as a power
law in the limit of $\varepsilon \rightarrow 0$,
\vspace*{-0.2cm}
\begin{equation}
\mu(\varepsilon) \approx \mu (0) + A \varepsilon^{\beta}\;,
\end{equation}
\vspace*{-0.08cm}
where $\mu (0)$ is the measure of chaotic parameters. $\beta$ was called the
fatness exponent. For quadratic maps it was found to be $\beta
\simeq 0.45$. Since our map belongs to the same universality class as
considered in Farmer's case, namely the map has a single quadratic maximum,
one may expect that $\beta$ will have the same value.  For the climbing sine
map a double-logarithmic plot of $\delta \mu (\varepsilon) = \mu(\varepsilon)
-\mu(0)$ is shown in Fig.\ \ref{Fig8} for the first three boxes, $j=0,1,2$.
In all cases the bold lines have the slope $0.45$ with errors of $0.03, 0.04,
0.05$, respectively, which seems to be in agreement with Farmer's conjecture
about the universality of $\beta$. However, apart from this coarse linear
behavior one can see an interesting oscillatory behavior in $\delta \mu$ with
respect to $\varepsilon$. This fine structure can be explained with respect to
the histogram distribution functions $f$ of the window sizes
$d$, which are plotted in Fig.\ \ref{Fig8} in form of dashed lines. Somewhat
surprisingly, the periodic windows are not distributed uniformly or smoothly
with respect to their size but form certain peaks, in which preferably windows
of certain periods are grouped together. This non-uniformity is clearly
reflected in the oscillations of $\delta\mu(\varepsilon)$. Moreover, we find
that the size distributions of periodic windows have a fine structure that
appears to resemble a fractal function. Some evidence for this property is
given in the inset of Fig.\ \ref{Fig8} (c), which shows the size of every window
of period five in the subset $j=2$ as a function of its appearance with
respect to the map control parameter $a$, i.e., not the parameter itself is
plotted but just an integer running index $N$ is given instead. Particularly
the height of the peaks is important, and one can clearly see a complicated
hierarchy of different peaks which are reminiscent of the fine structure in
the corresponding distribution function shown in Fig.\ \ref{Fig8} (c).

\section{Conclusions}
In this paper we have performed a detailed analysis of the parameter
dependence of the diffusion coefficient in a nonhyperbolic dynamical
system. The climbing sine map has been chosen as a paradigmatic example of
such a system. We have shown that, on a coarse scale, there are certain
analogies between the parameter-dependent diffusion coefficient of this map
and the ones in simple hyperbolic piecewise linear maps, such as the existence
of an oscillatory structure, and the existence of asymptotic functional forms
as derived from simple random walk models. However, in contrast to hyperbolic
maps showing normal diffusion only, in the nonhyperbolic climbing sine map
fractal structures appear for both normal and anomalous diffusive regions of
the diffusion coefficient. An understanding of the origin of these fractal
structures was given in terms of dynamical correlations starting from the
Green-Kubo formula for diffusion. We furthermore related these irregularities
in the diffusion coefficient more microscopically to different characteristics
in corresponding fractal generalized Takagi functions. For this purpose we
derived a new functional recursion relation that defines these fractal forms
and generalizes ordinary de Rham-type equations. Our analysis was completed by
extensive numerical studies of the periodic windows of the climbing sine map
showing that both the periodic and the chaotic parameter regions have positive
measures in the parameter space. However, these measures are themselves
parameter-dependent, and by increasing the parameter we found that the chaotic
regions occupy larger and larger measures. We finally provided evidence that
these different sets form fat fractals on the parameter axis.
  
In conclusion, we wish to remark that the climbing sine map is of the same
functional form as the respective nonlinear equation in the two-dimensional
standard map, which is considered to be a standard model for many physical
Hamiltonian dynamical systems. Indeed, both models are motivated by the driven
nonlinear pendulum, both are strongly nonhyperbolic, and though the standard
map is area-preserving it too exhibits a highly irregular parameter-dependent
diffusion coefficient. Understanding the origin of these irregularities was
the subject of intensive research \cite{Ott97,Rec80}, however, so far the
complexity of this system did not enable to reveal its possibly fractal
nature. A suitably adapted version of our approach to nonhyperbolic diffusive
dynamics as presented in this paper may enable to make some progress in this
direction.

Another interesting problem is to possibly further exploit the concept of
conjugacy between nonlinear and piecewise linear maps, as explained in Sec.\
III, in order to exactly calculate diffusion coefficients for nonlinear maps.
A very promising approach in this direction was presented in Ref.\
\cite{Alo96}. Based on these techniques we are planning to perform a spectral
analysis of the Frobenius-Perron operator governing the probability density of
the diffusive climbing sine map. Combining such an analysis with the Takagi
function approach outlined here may lead to a general theory of nonhyperbolic
transport.

It would furthermore be important to check out the applicability of periodic
orbit theory for computing the parameter-dependent diffusion coefficient of
the climbing sine map, which may provide an alternative method
\cite{Cvi01}. Another promising direction of future research concerns
establishing crosslinks between our work and the realm of strange kinetics and
stochastic modeling as described in Refs.\
\cite{Zum93,Klaf94/95,Met00}, e.g., by trying to apply continuous time random walk
techniques to more complicated chaotic models exhibiting fractal diffusion
coefficients such as the climbing sine map.

We finally emphasize the importance to look for possibly fractal transport
coefficients in experiments. A very promising candidate appears to be the
phase dynamics in SQUID's, which was very recently analyzed theoretically
\cite{TKK02} and studied experimentally \cite{Weiss}.

\appendix
\section{Diffusion coefficients of two conjugate maps}

In this Appendix we give a proof of the statement of Grossmann and Thomae
\cite{Gro83} that two diffusive maps which are conjugate
to each other have the same diffusion coefficient.

Two diffusive maps $F: I\to I$ and $G: J\to J$ are called conjugate
\cite{Gro77,Gro83,LM94} if there exists a map $H: I\to J$ such that $F(x) =
H(G(H^{-1}(x)))$.  Let us assume in the following that the conjugation
function $H$ is sufficiently smooth. Let the invariant densities of the
corresponding reduced ($mod \; 1$) maps be $\tilde{\rho} (x)$ for $F(x)
\; mod \; 1$ and $\rho (y)$ for $G(y) \; mod \; 1$; then it is, according to
conservation of probability, $ \tilde{\rho} (x) = |(H^{-1}(x))'| \; \rho
(H^{-1}(x)) $. The diffusion coefficients of the maps $F(x)$ and $G(y)$ we
denote by $D_F$ and $D_G$, respectively.  Without loss of generality let us
furthermore assume that the maxima of both maps are in the interval $[1, 2]$.

We now start with the Green-Kubo formula written in {\it correlated random
walk} terms as
\begin{equation}
 D_F   =  \frac{1}{2} \int\limits^1_0 \left[F(x)\right]^2 \tilde{\rho} (x) dx
+   \int\limits^1_0  \left[ F(x) \right] \cdot B(x) \; \tilde{\rho} (x) dx,
\label{ap1}
\end{equation}
where 
\begin{equation}
B(x) = \left[ F\{F(x)\}\right] + ... + \left[ F\left( \{ F \left( \{...\left(
\{ F(x) \} \right)...\} \right) \} \right)\right] + ...\;,
\label{ap2}
\end{equation}
or shortly
\begin{equation}
 D_F = d_{F_0} + d_{F_1} + d_{F_2} + ... 
\label{ap2b}
\end{equation}
where
\begin{equation}
d_{F_0} = \frac{1}{2}\int\limits^1_0 \left[F(x)\right]^2 \; \tilde{\rho} (x) \; dx,
\label{ap3}
\end{equation}
\begin{equation}
d_{F_1} = \frac{1}{2}\int\limits^1_0 \left[F(x)\right] \;
\left[F\{F(x)\}\right] \; \tilde{\rho} (x) \; dx,
\label{ap4}
\end{equation}
and so on. Focusing on the first term, one can rewrite this expression using
the symmetry of the map to
\begin{equation}
d_{F_0} = \frac{1}{2}\int\limits^1_0 \left[F(x)\right]^2 \; \tilde{\rho}(x) \; dx  =
 \int\limits^{x2}_{x1} \tilde{\rho}(x) \; dx,
\label{ap5}
\end{equation}
where $x_1, x_2$ defines an escape region. For the conjugate map $G(y)$ the
respective term reads
\begin{equation}
d_{G_0} = \frac{1}{2}\int\limits^1_0 \left[G(y)\right]^2 \; \rho(y) \; dy  =
 \int\limits^{y2}_{y1} \rho(y) \; dy, 
\label{ap6}
\end{equation}
where $y_1, y_2$ is the corresponding escape region for $G$. Note that the
escape regions $(x_1, x_2)$ and $(y_1, y_2)$ are not the same, however, it is
straightforward to show that $H(x_i)=y_i\:,\:i=1,2$, that is, the topology of
both maps is conserved such that the two escape regions are mapped onto each
other under conjugacy.

Taking into account the conservation of probability mentioned before one
immediately gets
\begin{equation}
d_{F_0} = d_{G_0}\;.
\label{ap7}
\end{equation}
All other terms $d_{F_1}, d_{F_2}, ...$ and $d_{G_1}, d_{G_2}, ...$ have the form
\begin{equation}
d_{(F,G)_i} = A \; \int\limits_{\delta_{esc}} \nu (z) \; dz, \; \; i = 1, 2, ...
\label{ap8}
\end{equation}
where $A$ is a constant, $\delta_{esc}$ is the respective escape region and
$\nu (z) \; dz$ is the corresponding invariant measure. Thus, the same
argument can be applied to show that $d_{F_i} = d_{G_i}$, ($i = 1, 2, ...$).
Combining all results we arrive at
\begin{equation}
 D_F = D_G \;.
\label{ap9}
\end{equation}
\section{Recursion relation for generalized Takagi functions}

We start with the recursion relation for the jump velocity function Eq.\
(\ref{rec}),
\begin{equation}
J_a^n (x) =  j_a(x) + J_a^{n-1} (m_a(x)),
\label{AppB_1}
\end{equation}
by recalling the definition of the generalized Takagi function Eq.\
(\ref{Takagi}),
\begin{equation}
T_a^n(x) := \int_0^x J_a^n (z) \; dz, \; \; \; T_a^n(0)\equiv
T_a^n(1) \equiv 0,
\label{AppB_2}
\end{equation}
or differently
\begin{equation}
J_a^n (x) = \frac{d}{dx} T_a^n(x).
\label{AppB_2_a}
\end{equation}
\vspace*{0.1cm}
\begin{figure}[htb]
\centering
\includegraphics[width=0.6\textwidth]{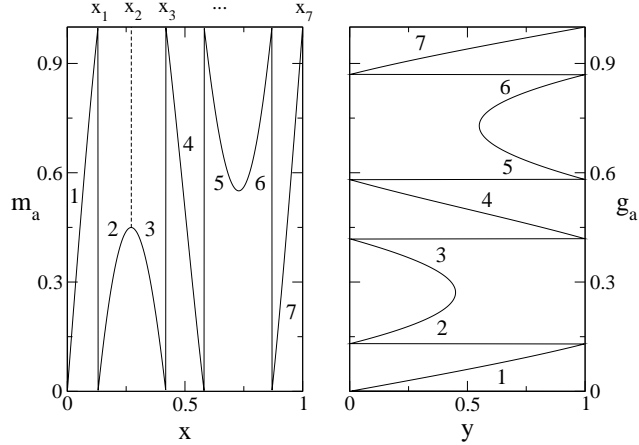}\\
\caption{Illustration of the construction of the inverse function of the
climbing sine map for the parameter value $a=1.189$. Piecewise invertible
branches are labeled by the integer numbers $i=1, \cdots, 7$.}%
\label{App_L}
\end{figure}

We have to integrate Eq.\ (\ref{AppB_1}),
\begin{equation}
\int\limits_{0}^{x} dy \; J_a^n (y) = \int\limits_{0}^{x} dy \; j_a(y) + 
\int\limits_{0}^{x} dy \; J_a^{n-1} (m_a(x)).
\label{AppB_3}
\end{equation}
By using of Eq.\ (\ref{AppB_2}) we get
\begin{equation}
T_a^n (x) =  t_a(x) + I(x), \; x\in (0,1],
\label{AppB_4}
\end{equation}
where
\begin{equation}
I(x) = \int\limits_{0}^{x} dy \; J_a^{n-1} (m_a(y)), \; x\in (0,1].
\label{AppB_5}
\end{equation}
Without loss of generality let us assume that the maximum of the map is in
the interval $[1, 2]$. 

Depending on $x$ the integral in Eq.\ (\ref{AppB_5}) can be decomposed into
\begin{equation}
I_1(x) = \int\limits_{0}^{x} dy \; J_a^{n-1} (m_a(y)), \; x \in (0,x_1];
\label{AppB_6}
\end{equation}
\begin{eqnarray*}
I_2(x) = \int\limits_{0}^{x_1} dy \; J_a^{n-1} (m_a(y)) \; + 
\int\limits_{x_1}^{x} dy \; J_a^{n-1} (m_a(y)), \; x \in (x_1,x_2]; 
\end{eqnarray*}
\[
I_3(x) = ..., \; x \in (x_2,x_3]; \; ... \; I_6(x) = ..., \; x \in (x_5,x_6];  
\]
\begin{eqnarray*}
I_7(x) = \int\limits_{0}^{x_1} dy \; J_a^{n-1} (m_a(y)) + \cdots 
+ \int\limits_{x_6}^{x} dy \; J_a^{n-1} (m_a(y)), \; x \in (x_6,x_7]. 
\end{eqnarray*}
Each integral in Eq.\ (\ref{AppB_6}) now contains only one piecewise
invertible branch of the reduced map $m_a^i(x)$ as shown in Fig.\
\ref{App_L}. Here, the piecewise invertible branches of the reduced map are labeled by
integers, and the corresponding branches of the inverse function $g_a^i(y)$
have the same indices $i=1, \cdots, 7$.  Since all integrals in Eq.\
(\ref{AppB_6}) have the same form (only the inverse parts of the reduced map
are different), we restrict ourselves to the integral
\begin{equation}
I(x) = \int\limits_{0}^{x} dy \; J_a^{n-1} (m_a^i(y)).
\label{AppB_7}
\end{equation}
Making the change of variables $z=m_a^i(y)$ and using the definition of the
generalized Takagi function Eq.\ (\ref{AppB_2_a}) we get
\begin{equation}
I(x) = \int\limits_{0}^{m_a^i(x)} dz \; (g_a^i(z))^{'} \frac{d}{dz} T_a^{n-1}(z).
\label{AppB_8}
\end{equation}
Using integration by parts we arrive at
\begin{equation}
I(x) = (g_a^i(z))^{'} \cdot T_a^{n-1}(z)|_{0}^{m_a^i(x)} - 
\int\limits_{0}^{m_a^i(x)} dz \; (g_a^i(z))^{''} \cdot T_a^{n-1}(z).
\label{AppB_9}
\end{equation}
Now recall that according to Eq.\ (\ref{AppB_2}) it is
$T_a^{n-1}(m_a^i(x_j)) \equiv 0$, where the $x_j$ define the boundaries of the
piecewise invertible parts of $m_a(x)$, see Fig.\ \ref{App_L}, and that
$(g_a^i(z))^{'}|_{z=m_a^i(x)}\equiv 1/(m_a^i(x))^{'}$.  Thus, by formally
defining the inverse function $g_a(x)$ as consisting of all branches
$i=1,\ldots,7$, we can finally write Eq.\ (\ref{AppB_4}) in the form
\begin{equation}
T_a^{n}(x) = t_a (x) + \frac{1}{m^{'}_{a}(x)} \;
T_a^{n-1} (m_a(x)) - I(x)
\label{AppB_last}
\end{equation}
with the integral term
\begin{equation}
I(x) = \int_{0}^{m_a(x)} dz g_a''(z) T_a^{n-1} (z).
\label{AppB_last_1}
\end{equation}
\section{Numerical procedure for calculating the measure of the periodic windows}
The parameter values $a_{tan}$ which correspond to the tangent bifurcations of
the $p$-periodic windows were found by solving the two coupled
transcendental equations
\begin{equation}
\label{tanbif}
\partial m_{a_{tan}}^{(p)} (x) / \partial x =  1, \; m_{a_{tan}}^{(p)} (x) - x = 0,
\end{equation}
where $m_{a}^{(p)}(x)$ denotes the $p$-times iterated reduced map.  This
corresponds to the situation where $m_{a}^{(p)}(x)$ touches the bisector.
Somewhat after a tangent bifurcation one will unavoidably find a situation
where a critical point $x_{c}$, which corresponds to an extremum of
$m_{a}^{(p)}(x)$, crosses the bisector. When this critical point is exactly
located on the diagonal, the reduced map or its higher iterations have a fixed
point and there exists a specific Markov partition on the interval
\cite{Kla95,Kla96}. The periodic orbit generated by the corresponding parameter
value $a_{ss}$ is superstable,
\begin{equation}
\label{sstable}
m_{a_{ss}}^{(p)} (x_c) - x_c = 0.
\end{equation} 
By further increasing the parameter value up to $a_{cr}$ a crisis takes place,
and this again corresponds to the existence of a certain Markov partition.

Based on this scenario, the full numerical procedure which was used for
calculating the measure of periodic windows is as follows: The values of
$a_{ss}$ corresponding to superstable solutions were first calculated by a
combination of bisection with the Newton method.  The parameters for the
tangent bifurcations could then usually be found by the modified
two-dimensional Newton method \cite{Far85}. However, the highly discontinuous
nature of $m_{a}^{(p)}(x)$ made its implementation very inefficient.  Instead,
starting in the vicinity of each $a_{ss}$ we again combined the
one-dimensional Newton and bisection methods. This ensured that no windows
were missed.  Finally, the parameter values corresponding to crisis points
$a_{cr}$, which are also defined by Markov partitions, can be found by 
solving respective equations that are formally analogous to
Eq.\ (\ref{sstable}).

\end{document}